\documentstyle[12pt]{article}

\newcommand{\be}{\begin{equation}}
\newcommand{\ee}{\end{equation}}

\begin{document}

\bibliographystyle{unsrt}
\footskip 1.0cm
\thispagestyle{empty}
\setcounter{page}{0}
\begin{flushright}
IC/95/220\\

August 1995\\
\end{flushright}
\vspace{10mm}

\centerline {\LARGE CAN SYMMETRY NON-RESTORATION}
\vspace{5mm}
\centerline {\LARGE SOLVE THE MONOPOLE PROBLEM ?}
\vspace*{15mm}
\centerline {\large G. Bimonte}
\vspace{5mm}
\centerline{ \small and }
\vspace*{5mm}
\centerline{ \large G. Lozano \footnote{
E-mail addresses: bimonte@napoli.infn.it~~,~~lozano@ictp.trieste.it}}
\vspace*{5mm}
\centerline {\it International Centre for Theoretical Physics, P.O.BOX 586}
\centerline {\it I-34100 Trieste, ITALY}
\vspace*{25mm}
%\baselinestretch{2.0}
\normalsize
\centerline {\bf Abstract}
\vspace*{5mm}
{\large We reexamine a recently proposed non-inflationary solution to the
monopole problem, based on the possibility that spontaneously broken
Grand-Unified symmetries do not get restored at high temperature. We go
beyond leading order by
studying the self-consistent one-loop equations of the model. We find
large next-to-leading corrections that reverse the lowest order results
and cause symmetry restoration at high temperature. } \newpage

\baselineskip=24pt
\setcounter{page}{1}

\section{Introduction}

Since the pioneering works by Kirzhnits and Linde \cite{lin}, Weinberg
\cite{wei} and Dolan
and Jackiw \cite{dol}, the analysis of the phase diagrams of gauge theories
at finite temperature has been object of intense research. Many of these
studies are connected with spontaneously broken gauge theories where the
typical picture which emerges is that symmetries get restored at high
temperatures.

In standard early universe scenarios based on Grand Unified Theories
(GUT's), one then starts from the {\it symmetric} phase of the model, and
as the temperature is lowered, several phase transitions occur until the
broken phase described by the Standard Model is reached.
A generic feature of this process is the unavoidable formation of
topological defects during these phase transitions and, as it is well known,
the over abundant production of monopoles at this stage represents one of
the most serious drawbacks of GUT's.

Different solutions to the monopole
problem
have been suggested along the years. Apart from inflation, which is the
most popular one, several authors have proposed alternative scenarios.
Among them, let us mention the work of Langacker and Pi \cite{lan}, who
have suggested that the existence of an intermediate phase, with no
unbroken $U(1)$ symmetry, would cause a rapid annihilation of the
monopoles produced during the GUT phase transition. Another and more
radical possibility has been considered by Salomonson, Skagerstam
and Stern \cite{sal} and very recently by
Dvali, Melfo and Senjanovi\'{c}
\cite{goran}. It is based on the
attractive idea that the monopole producing phase transition could have
not occurred at all. In other words, it is argued that, no matter how
high the temperature was, the symmetric phase of $SU(5)$ was never realized.

Although not very popular and somehow counter-intuitive, the phenomenon of
symmetry
non-restoration has been present in the literature for quite a while. In
fact, it dates back to the classic paper of Weinberg \cite{wei} and some
interesting
phenomenological applications have been discussed in the past \cite{ms}.

The possibility of symmetry non-restoration in GUT's is analyzed in
ref.\cite{sal} for the minimal $SU(5)$ model, which has two Higgs
fields in the representations $\bf 5$ and ${\bf 24}$. Neglecting the
gauge couplings, it was shown there that there exists a range of parameters
for the Higgs potential, leading to the  symmetry of the Standard Model
at low energies but such that the heavy Higgs responsible of the $SU(5)$
breaking keeps a non vanishing vev even at very high temperatures.  Gauge
couplings are considered, instead, in
ref.\cite{goran}, where it is shown that, due to the large value of the
gauge-coupling constant at the scale of the $SU(5)$ phase transition,
symmetry non-restoration in the
minimal model requires such large values for the Higgs parameters that
perturbation theory breaks down. It is subsequently observed that
if the Higgs sector is enlarged with
a scalar in ${\bf 45}$ representation of $SU(5)$
symmetry non-restoration can be achieved for
values of the Higgs coupling constants which seem
small enough as to make reliable the lowest order
computation on which the whole analysis is based.

In this paper, we will reanalyze the phenomenon of symmetry
non-restoration for the GUT models considered in ref.\cite{goran} by
taking into account next-to-leading order corrections. The reasons for
doing this stem from our study of the simpler $O(N_1)\times O(N_2)$
scalar model \cite{bim} (which mimics the scalar sector of GUT's) where
we found that the inclusion of sub-leading corrections reduces in a
significant way the region of parameter space where symmetry
non-restoration occurs. As we will show in this paper, the effect of
including these corrections is even more dramatic here as
they reverse the lowest order result and cause symmetry restoration
in the entire range of parameters that was considered in ref.\cite{goran}.

The paper is organized as follows. In Section 2 we present the model and
discuss the phenomenon of symmetry non restoration to lowest order, while
in Section 3 we present the results of our next-to-leading calculations.
Finally, we leave for Section 4 some final remarks.

\section{A review of symmetry non-restoration}

In this Section, we shall briefly review the basic results about symmetry
non-restoration in lowest order perturbation theory. For definiteness, we
shall consider the GUT model of ref.\cite{goran},
having the usual fermionic
content, namely, three generations of fermions $(\Psi_L^f)_{5^*}$,
$(\Psi_L^f)_{10}$, $f=1,2,3$ in the
${\bf 5^*}$ and ${\bf 10}$ representations respectively and three Higgs
multiplets, $\Phi_{5}$, $H_{24}$ and
$\chi_{45}$ in the complex
${\bf 5}$,
real ${\bf 24}$ and complex ${\bf 45}$ dimensional representations
respectively.

The model is described by the Lagrangian
$$
L= -\frac{Tr}{2} F_{\mu \nu} F^{\mu \nu} +
 \frac{1}{2}(D_{\mu}H)_i (D^{\mu}H)_i + (D_{\mu}\Phi)^*_a(D^{\mu}\Phi)_a
+ (D_{\mu}\chi)_u^* (D^{\mu}{\chi)_u}+
$$
\be
 - V(H_i,\Phi_a,\chi_u) + L_{F}~~,
\ee
where, in order to simplify the notation, we have called $\Phi_a,
a=1,\cdots 5~$, $H_i, i=1,..,24~$
and $\chi_u, u=1,...,45$ the independent components of the Higgs fields
$\Phi_{5}$, $H_{24}$ and $\chi_{45}$ respectively \footnote{Here and
in what follows we shall adopt the following indices convention: Greek
letters will be used for space-time indices, while Latin letters will
denote group indices. Group indices will be denoted
with the initial letters of the alphabet $a,b,c...$ when running from 1
to 5, with the middle letters $i,j,k...$ when running from 1 to 24 and with
the last letters $u,v,w...$ when running from from 1 to 45} and
\be F_{\mu \nu}=( \partial_{\mu}A^i_{\nu} - \partial_{\nu}A^i_{\mu}- g
f^{ijk}A_{\mu}^j A_{\nu}^k)T^i~,
\ee
\be
[T^i,T^j]=if^{ijk}T^k~,~~~~~Tr(T^iT^j)=\frac{1}{2}\delta^{ij}~~,\label{com}
\ee
\be
(D_{\mu}\Phi)_a=\partial_{\mu}\Phi_a-igA^j_{\mu}(T^j)_{ab} \Phi_b~.
\ee
\be
(D_{\mu}H)_i=\partial_{\mu}H_i-igA^j_{\mu}(T^j_H)_{ik}H_k~,
\ee
\be
(D_{\mu}\chi)_u=\partial_{\mu}\chi_u-igA^j_{\mu}(T^j_{\chi})_{uv}\chi_v~.
\ee
Here, $T^j_H,~T^j_{\chi}$ are the $SU(5)$ generators in the ${\bf 24}$ and
${\bf 45}$ representations, satisfying the same commutation relations
as in eq.(\ref{com}) and such that:
\be
Tr(T^i_H T^j_H)=c_H \delta^{ij}~~,
\ee
\be
Tr(T^i_{\chi}T^j_{\chi})=c_{\chi}\delta^{ij}~,
\ee
where $c_H=5$ and $c_{\chi}=12$.
$L_F$ stands
for the fermionic Lagrangian, which we shall not write down explicitly
as it will play a minor role in our discussion.

As it is well known, the starting $SU(5)$ symmetry of the model can be
broken down to the low-energy standard model $SU(3)_c \times U(1)_{EM}$
symmetry in two stages. In the first, the field $H_{24}$ gets a vev
$\langle H \rangle$ that breaks $SU(5)$ to $SU(3)_c \times SU(2)_L
\times U(1)_Y$. In the second, the light fields $\Phi_5$ and/or
$\chi_{24}$ break this intermediate gauge symmetry down to $SU(3)_c
\times U(1)_{EM}$. It is during the first of these phase transitions that
monopoles form with the Kibble mechanism. As we explained in the
introduction, we are studying here the
possibility that the phase transition leading to the formation of
monopoles does not occur, namely that the heavy field $H_{24}$ keeps a
non-vanishing vev at
arbitrarily large temperatures.

According to the mechanism discussed by
Weinberg \cite{wei}, this goal can be achieved if one lets the field
$H_{24}$
interact, with a negative coupling, with the other Higgs fields $\Phi_5$
and/or $\chi_{24}$. Such couplings are naturally allowed in the model we are
considering. If one writes the most general SU(5)
invariant renormalizable potential \footnote{We are using a normalization
for the Higgs fields and coupling constants different from that of
ref.\cite{goran}. The relation between the two is as follows: $m^2_{\chi}=
2m^{\prime 2}_{\chi},~m^2_H=4m^{\prime 2}_H,
{}~\lambda_{\chi}=4\lambda^{\prime}_{\chi},~\lambda_H=16
\lambda^{\prime}_H,~\alpha=4 \alpha^{\prime}$, where the primed constants
correspond to those used in \cite{goran}.}
as: $$ V(H_i,\Phi_a,\chi_u)= -\frac{1}{2} m_H^2 H_iH_i
- m_{\Phi}^2 \Phi_a^*\Phi_a- m_{\chi}^2\chi_u^*\chi_u
+ \lambda_{\chi} (\chi_u^* \chi_u)^2 +
$$
\be
+ \frac{1}{4}\lambda_H (H_i H_i)^2
+\lambda_{\Phi}(\Phi_a^*\Phi_a)^2
- \alpha (\chi_u^* \chi_u) (H_i H_i)+v(H_i,\Phi_a,\chi_u)~,\label{pot}
\ee
it is for example possible to obtain the desired symmetry breaking pattern
at zero temperature with a $v(H_i,\Phi_a,\chi_u)~
$ negligible compared to the other terms.
Then the condition of boundedness from below on $V$ reduces to:
\be
\lambda_{\chi} \lambda_H > \alpha ^2,~~~~\lambda_{\chi}>0,~~\lambda_H>0~.
\label{bcon}
\ee
The sign of $\alpha$ is thus unconstrained and one can choose it positive
such as to give a negative interaction between $H_{24}$ and $\chi_{45}$.
Obviously
one may also consider the analogous case where $\chi_{45}$ is replaced by
$\Phi_5$
in the $\alpha$ dependent interaction term in eq.(\ref{pot})\cite{sal}.
These are
precisely the two cases considered in ref.\cite{goran}, and on which we
will concentrate our attention here.

Let us briefly review how symmetry non-restoration arises in the model.
As it is well known, the basic tool to study the symmetry behavior of a
theory at finite temperature is provided by the effective potential
\cite{dol}. As we are only interested here in establishing the existence
(or not) of a stable symmetric phase at arbitrarily high temperatures
($T~ \gg \langle\Phi_a\rangle,~\langle H_i \rangle,~
\langle \chi_u \rangle$) and not in the details of the phase transition, we
will only need to
compute the leading terms, of order $T^2$, of the effective potential. This
turns out to be equivalent to evaluating the leading $T^2-$contributions
to the self-energies of the Higgs fields (Debye masses).

The result for the Debye masses to lowest order in the
coupling constants can be obtained by adapting Weinberg's formulae to
this model and read \cite{kuz}:
\be m^2_{\chi}(T)=\Sigma_{\chi}(T,p=0)=
T^2\left(\frac{1}{6}\lambda_{\chi}(1+N_{\chi})-\alpha
\frac{N_H}{12}+\frac{1}{4}g^2 \frac{D c_{\chi}}{N_{\chi}}\right)
\equiv \nu^2_{\chi}T^2~~,\label{low1}
\ee
\be
m^2_{H}(T)=\Sigma_{H}(T,p=0)=
T^2\left(\frac{1}{12}\lambda_{H}(2+N_{H})-\alpha
\frac{N_{\chi}}{6}+\frac{1}{4}g^2 \frac{D c_H}{N_H}\right)
\equiv \nu^2_H T^2~~,\label{low2}
\ee
where $N_H=24$ and $N_{\chi}=45$ are the dimensions of the
representations of the Higgs fields and $D=24$ is the dimension of the
group. The different numerical factors between
eqs.(\ref{low1}) and (\ref{low2}) arise from the fact that $\chi_{45}$ is
complex while $H_{24}$ is real (the analogous result for the case of a
coupling between $H_{24}$ and $\Phi_5$ can be obtained by simply
substituting in eqs.(\ref{low1}) and (\ref{low2}) to $N_{\chi}$ and
$c_{\chi}$ the corresponding quantities $N_{\Phi}$ and $c_{\Phi}$).

These formulae exhibit the basic idea behind symmetry non-restoration: for
positive $\alpha$'s
the interaction $H_{24}-\chi_{45}$ term
gives a negative contribution to the thermal masses and might make one of
them negative (it is easy to see that condition (\ref{bcon}) implies that
they cannot be simultaneously negative). Once this happens the
corresponding Higgs field maintains a
vev at high temperatures and symmetry is never restored.

As explained
earlier, we wish to have $\langle H_{24} \rangle \neq 0$ at high $T$ and
according to eq.(\ref{low2}) this requires:
\be
m^2_H(T)<0 ~~\Rightarrow ~~\alpha >
\lambda_H \frac{2+N_H}{2N_{\chi}}+
\frac{3}{2}g^2 \frac{D c_H}{N_H N_{\chi}}~~\label{bre}.
\ee
The set of coupling constants satisfying this inequality, together with
the boundedness condition, eq.(\ref{bcon}), thus represents, to lowest order,
the
region of parameter space in which symmetry non-restoration occurs.

The presence of gauge couplings plays an essential role: as it is evident
from eqs.(\ref{low1}) and (\ref{low2}), it conspires against symmetry
non-restoration. As a consequence, $\alpha$ has to be large enough to
overcome their contributions to the thermal masses. But, on the other
side, if $\alpha$ becomes too large,
as a consequence of the boundedness condition eq.(\ref{bcon})
$~\lambda_{\chi}$ or $\lambda_H$ are pushed outside the range of
applicability of perturbation theory and the whole analysis breaks down. In
fact, combining eq.(\ref{bre}) with eq.(\ref{bcon}) one gets:
\be
\lambda_{\chi}>\frac{1}{\lambda_{H}}\left(
\lambda_H \frac{2+N_H}{2N_{\chi}}+
\frac{3}{2}g^2 \frac{D c_H}{N_H N_{\chi}}\right)^2~~.
\ee
The r.h.s. has a minimum for
\be
\hat{\lambda}_H=3 g^2 \frac{D c_H}{N_H(N_H+2)}~,
\ee
which gives a lower bound for $\lambda_{\chi}$ and $\alpha$:
\be
{\lambda}_{\chi}> \hat{\lambda}_{\chi}= \hat{\lambda}_H\left(
\frac{N_H+2}{N_{\chi}}\right)^2~~,\label{min1}
\ee
\be
\alpha > \hat{\alpha}=\hat{\lambda}_H \frac{2+N_H}{N_{\chi}}.\label{min2}
\ee
Eqs.(\ref{min1})(\ref{min2}) make it apparent that there are better
chances of keeping $H_{24}$ broken at all temperatures with small
values of the coupling constants by coupling it with
Higgses belonging to representations
of large dimensions. This remark is relevant because of the large value of
$g^2$, tipycally taken to be $g^2 \approx 1/4$. With this value for
$g^2$, one has in fact:
\be
\hat{\lambda}_H \approx 0.15~~,\label{cons1}
\ee
\be
\hat{\lambda}_{\chi}\approx 0.05~~,\label{cons2}
\ee
\be
\hat{\alpha}\approx 0.09~~.\label{cons3}
\ee
Had one coupled instead $H_{24}$ to $\Phi_5$, one would have found:
\be
\hat{\lambda}_{\Phi}\approx 3.9~~,
\ee
\be
\hat{\alpha}\approx 0.75~~,
\ee
While it is more or less clear that for the ${\bf 5}$ representation one
needs
too large values of the coupling constants for perturbation theory to be
reliable, there seems to be some hope with the $\bf 45$ representation.
Nevertheless the reader should be aware that loop corrections may contain
powers of $N_{\chi}\lambda_{\chi}$ and thus also for the case of a
coupling between $H_{24}$ and $\chi_{45}$
an analysis of higher order corrections results unavoidable.

\section{Next-to-leading order corrections}

In this Section  we will compute
the next to leading order
corrections to the thermal masses of the Higgs fields to determine if the
results of the lowest order analysis of the previous Section are
reliable, when the coupling constants are of the order of those of
eqs.(\ref{cons1}-\ref{cons3}).

At this point, let us mention that some authors \cite{fuj} have argued
that Weinberg's results on symmetry non-restoration are only an artifact
of perturbation theory and that in reality symmetry is always
restored a high temperatures. Their claims are based on the observation
that non-perturbative calculations (such as $1/N$ expansion and Gaussian
Effective Potential) when applied to $O(N_1)\times O(N_2)$ global scalar
models give always symmetry restoration. Even though non perturbative,
the methods used in \cite{fuj} are approximations, whose range of validity
are not clear to us. We will follow a more standard route, by assuming
that perturbation theory (conveniently improved) is valid and determining
whether or not the lowest order results remain true after
next-to-leading order corrections are included.

Due
to the infrared characteristic behavior of field theories at high
temperatures,
the corrections to the self-energies are expected to be of order $e^{3/2}$
instead of
$e^2$, where $e$ is the largest among $\lambda_{\chi}, \lambda_{H},
\alpha, g^2$. A simple way of obtaining these corrections is via the
analysis of the so-called ``gap equations", which correspond to a
one-loop truncation
of the Schwinger-Dyson equations for the self-energy and which are
equivalent to
a resummation of the ring diagrams of the perturbative series.
Alternatively,
they can also be derived by adding and subtracting from the Lagrangian a
temperature-dependent mass counter term \cite{par}, which is then determined
self-consistently in such a way as to cancel $T^2$-divergent terms occurring
in the
self-energies. This kind of improvement of perturbation theory has been
discussed recently in the context of studies on the Electroweak phase
transition \cite{qui}.
We will write the gap equations equations in the symmetric  phase and we
will identify
the region of parameter space for which real solutions for the masses
can be found as the region of symmetry restoration.

The one-loop gap equations we are considering are schematically represented
in fig.(1). Notice that we have neglected the Yukawa couplings between
the Higgs fields and the fermions. We will comment on this point below.
The blobs in the internal lines of the diagrams shown there
represent the complete thermal propagators of the fields. To the order we
are interested in, they can be approximated with free-like propagators
containing the thermal masses instead of the zero-temperature ones. For the
Higgs fields this corresponds to using (in Euclidean space-time):
\be
\langle H_i(-p)H_j(p)\rangle= \frac{\delta_{ij}}{p^2+m^2_H(T)}~~,
\ee
\be
\langle \chi_u(-p)\chi_v(p)\rangle= \frac{\delta_{uv}}{p^2+m^2_{\chi}(T)}~.
\ee
As for the gauge fields $A_{\mu}^i$, their propagators are written in the
Landau gauge and read:
\be
\langle A^i_{\mu}(-p)A^j_{\nu}(p)\rangle=
\frac{\delta^{ij}}{p^2+m^2_L(T)}(P_L)_{\mu \nu}+
\frac{\delta^{ij}}{p^2}(P_T)_{\mu \nu}~~,
\ee
where
\be
(P_T)_{\mu\nu}=\delta_{\mu r}\left(\delta_{rs} -
\frac{k_rk_s}{\vec{k}^2}\right) \delta_{s \nu}~.\label{PT}
\ee
and
\be
(P_L)_{\mu \nu}= \delta_{\mu \nu}- \frac{k_{\mu} k_{\nu}}{k^2} -
(P_T)_{\mu \nu}~.
\ee
(in eq.({\ref{PT}) $r$ and $s$ denote space indices $r,s=1,2,3$). Here
$m_L(T)$ stands for the longitudinal thermal mass of the gauge bosons.
The standard high-temperature one-loop result for $m_L(T)$ in an $SU(N)$
gauge theory with $F_i$ chiral fermions, $h_j$ real Higgs and $\Sigma_k$
complex Higgs in the representations $R_i,~ R_j,~ R_k$ respectively is
given by:
\be
\nu^2_g \equiv \frac{m^2_L(T)}{T^2}=g^2\left(\frac{N}{3}+
\sum_{F_i}\frac{c_{R_i}}{6}+\sum_{h_j}\frac{c_{R_j}}{6}+
\sum_{\Sigma_k}\frac{c_{R_k}}{3}\right)~~.
\ee
Specializing this formula to our case, we get:
\be
\frac{m^2_L(T)}{T^2}= g^2\left( \frac{5}{3}+\frac{1}{2}c_{\bf 5}+
\frac{1}{2}c_{\bf 10}+ \frac{1}{6}c_H + \frac{1}{3}c_{\bf \Phi} +
\frac{1}{3}c_{\chi}\right)=\frac{23}{3}g^2~.\label{gau}
\ee
(Remember that we have three generations of fermions and that $c_{\bf
5}=1/2,~~c_{\bf 10}=3/2$.) In our calculation to order
$e^{3/2}$ this approximation for $m_L(T)$ is good enough.

The gap equations determining $m_{\chi}(T)$ and $m_H(T)$ to order
$e^{3/2}$ can
be obtained by evaluating the diagrams of fig.(1) at zero external
momenta. A simple computation gives the high-$T$ result:
\be
\left(\Sigma_{\chi}\right)_{uv}(p=0,T)=\left(
\Sigma_{\chi}^{a_1}+
\Sigma_{\chi}^{a_2}+
\Sigma_{\chi}^{a_3}+
\Sigma_{\chi}^{a_4}\right)\delta_{uv}~,\label{sig}
\ee
where
\begin{eqnarray}
\Sigma_{\chi}^{a_1} & = & 4 \lambda_{\chi}(1+N_{\chi})T^2
h\left(\frac{m_{\chi}^2(T)}{T^2}\right)~,\\
\Sigma_{\chi}^{a_2} & = & -2\alpha N_H T^2
h\left(\frac{m^2_H(T)}{T^2}\right)~,\\
\Sigma_{\chi}^{a_3} & = & 2 g^2 \frac{D c_{\chi}}{N_{\chi}}T^2\left[
h\left(\frac{m^2_L(T)}{T^2}\right)+2 h(0)\right]~,\\
\Sigma_{\chi}^{a_4} & = & 0~,
\end{eqnarray}
with a similar result for $(\Sigma_H)_{ij}(p=0,T)$. Here $h(y)$
represents the function:
\be
h(y^2)=\frac{1}{4\pi^2}\int_0^{\infty}dx
\frac{x^2}{(x^2+y^2)^{1/2}(e^{(x^2+y^2)^{1/2}}-1)}~,
\ee
which, for small values of $y$ has the asymptotic expansion
\be
h(y^2)=\frac{1}{24}-\frac{1}{8 \pi}\sqrt{y^2}-\frac{1}{16
\pi^2}y^2\left( \ln \frac{y^2}{8\pi}+\gamma -\frac{1}{2}\right)+\cdots~.
\ee
Upon using this expansion in eq.(\ref{sig}) and retaining up to the terms
linear in \\
$x_{\chi}\equiv \sqrt{m^2_{\chi}(T)/T^2}$ and $~x_{H}\equiv
\sqrt{m_{H}^2(T)/T^2}$ one gets the following gap equations:
\begin{eqnarray}
x_{\chi}^2 & = & \nu^2_{\chi} - \frac{g^2}{4 \pi}\frac{Dc_{\chi}}
{N_{\chi}}\nu_g
-\lambda_{\chi} \frac{1+N_{\chi}}{2 \pi} x_{\chi}
+\alpha \frac{N_H}{4\pi} x_H~,  \label{gap1}\\
x_{H}^2 & = & \nu^2_{H} - \frac{g^2}{4 \pi}\frac{D
c_{H}}{N_H}\nu_g
-\lambda_{H} \frac{2+N_H}{4 \pi} x_{H}
+\alpha \frac{N_{\chi}}{2\pi} x_{\chi}~~,\label{gap2}
\end{eqnarray}
where
$\nu_{\chi},~\nu_H$ are the same as in eqs.(\ref{low1})(\ref{low2}).
(including higher terms in the expansion would lead to corrections to the
thermal masses of order $e^2 \log e$, but then one would have to
consider also two-loops contributions to the Schwinger-Dyson equations,
which are of this order and that we have neglected in our one-loop
computation).

To lowest order eqs.(\ref{gap1}-\ref{gap2}) reproduce the results of
eqs.(\ref{low1})(\ref{low2}) for the thermal masses, but the inclusion of
the
next-to-leading corrections modifies the equations in an important way as
they
introduce a coupling among the thermal masses of the Higgs fields. Notice
that this coupling, represented by the term proportional to $\alpha$,
comes with a positive sign and then works against symmetry non-restoration.

Now, eqs.(\ref{gap1}-\ref{gap2}) represent a set of two parabolae
$x_H=P_H(x_{\chi}),~~x_{\chi}=P_{\chi}(x_H)~$ in the
$x_{\chi},~x_H$ plane.
When these parabolae intersect in the upper right part of this plane
(remember that $x_H$ and $x_{\chi}$ are defined to be positive), the
mass terms of both Higgs fields have a positive  sign at high temperature
and thus symmetry is restored.
The condition for the existence of such an
intersection is:
\be
P_{\chi}(x_H=0)<x^+_{\chi}
\ee
where $x_{\chi}^+$ is the positive root of the equation $P_H(y)=0$. After
some trivial algebra, this condition is found to be equivalent to
\be
\alpha <
f(g^2,\lambda_{\chi},\lambda_H,\alpha)~~,\label{ine}
\ee
where
$$
f(g^2,\lambda_{\chi},\lambda_H,\alpha) \equiv
\frac{2+N_H}{2N_{\chi}}\lambda_H + \frac{3}{2N_{\chi}}\left[ g^2
\frac{D c_H}{N_H}\left(1-\frac{\nu_g}{\pi}\right) \right]+
$$
\be
+\frac{3}{\pi}\alpha\left\{-
\lambda_{\chi}\frac{1+N_{\chi}}{4 \pi}+
\left[\left(\lambda_{\chi}\frac{1+N_{\chi}}{4 \pi}\right)^2+\nu^2_{\chi}
-g^2\frac{D c_{\chi}}{4 \pi
N_{\chi}}\nu_g\right]^{1/2}\right\}~.
\ee

The effect of including next-to-leading corrections can be better visualized
by fixing the coupling constant $\lambda_{\chi}$ and
showing how the region of symmetry non-restoration in the $(\lambda_H,
\alpha)$ plane changes in comparison to the lowest order result. To
lowest order, the values of coupling constants for which symmetry
non-restoration occurs are those in the region enclosed by the curve $c_1$
which represents the bound (\ref{bcon}):
\be
c_1~:~~\alpha=\sqrt{\lambda_{\chi} \lambda_H}
\ee
and the curve $c_2$ which represents the condition $m^2_H(T)=0$, see
eq.(\ref{bre}):
\be
c_2~~ : ~~\alpha=\lambda_H \frac{2+N_H}{2N_{\chi}}+
\frac{3}{2}g^2 \frac{D c_H}{N_H N_{\chi}}~~.
\ee
To next-to-leading order, the curve $c_2$ is shifted to the curve $c_3$
which is obtained by solving with respect to $\alpha$ the equation:
\be
c_3~~:~~\alpha=f(g^2,\lambda_{\chi},\lambda_H,\alpha)~.
\ee
The choice of $\lambda_{\chi}$ is not completely free, since it has to be
compatible with the bounds that come from an analysis of monopole
production in charged particles collisions \cite{tur}. In ref.\cite{goran}
it is shown that in order to fulfill this constraint one needs:
\be
\lambda_{\chi}>\frac{71}{135}g^2
\ee
which is a stronger bound than that given in eq.(\ref{min1}).
Figure (2), shows the curves $c_1,~c_2,~c_3$ for $g^2=1/4$ and
$\lambda_{\chi}=71 g^2/135$.
We see that the curve $c_3$ lies entirely above $c_1$,
which means that for all values of coupling constants compatible with the
bound (\ref{bcon}) there is symmetry restoration.

For completeness, we have considered also other values of
$\lambda_{\chi}$ and $g^2$. Keeping fixed $g^2=1/4$, we have found that
symmetry restoration occurs for all values of $\lambda_{\chi}$ consistent
with the bound (\ref{min1}). The results are instead more sensitive to the
choice of $g^2$. As it is clear already from the lowest order result and
our previous work on the ungauged model \cite{bim}, smaller values of
$g^2$ make symmetry non-restoration easier. As an indication, Figure (3)
shows the situation when $g^2=1/16$ and $\lambda_{\chi}=71/135 g^2$. We see
that even with this unrealistically small value of $g^2$ symmetry
non-restoration survives only in a tiny fraction of the lowest order region.
On the other hand, even this region disappears for larger values of
$\lambda_{\chi}$.

As we said earlier, in deriving our gap equations we neglected the Yukawa
couplings between the Higgs fields and the fermions. However, as they
always give a positive contribution to the Higgs self-energies, they
conspire against symmetry non-restoration and so can only strengthen our
conclusions.

\section{Conclusions}

In this paper we have analyzed the phenomenon of symmetry non-restoration
in a Grand Unified Model, beyond leading order. We have used linearized
gap equations to determine $e^3$ corrections to the thermal masses
of the Higgs fields at high temperature.
For realistic values of the $SU(5)$ gauge coupling
constant we
have found that next-to-leading order corrections to the thermal masses of
the Higgs fields completely
overwhelm the lowest order result and cause symmetry restoration.

An interesting question would of course be to determine the critical
temperature and the nature of the phase transition in this region of the
parameter space. To address this issue one should compute the
finite-temperature effective potential to an order consistent with the
linearized gap equations. As it is known \cite{qui,ame} a two-loop
computation
in the broken phase of the model is needed to perform this task. Notice
that also the form of the linearized gap equations would change, due to
the presence of additional couplings and to the fact that the
zero-temperature masses of the fields could not be neglected.

We have limited our analysis to the particular potentials
that were
considered in ref.\cite{goran}, which are clearly not the most general
$SU(5)$ invariant renormalizable potentials. To
our knowledge, the region of symmetry non-restoration has not been
analyzed for the general case even to lowest order.
In view of our results, which show that the mechanism
of symmetry non-restoration does not work in the simple models discussed
in \cite{goran},
an analysis of the general case
appears now to be necessary, before one can claim that
this mechanism can provide a solution to the monopole problem.
This analysis, although conceptually simple, becomes
tedious due to the large  number of free parameters (consider that the
general potential, even in the case of the model containing only
$H_{24}$ and $\chi_{45}$ depends on 16 real independent parameters).
Moreover, as our work shows, it will have to take into
account also the effects of subleading corrections.

In conclusion, we believe that the issue of whether the monopole problem
can be solved in realistic GUT
models  via the mechanism of symmetry non-restoration remains an open and
certainly interesting problem.

\section{Acknowledgements}

We would like to thank G. Senjanovi\'{c} and A. Melfo for the numerous
useful
discussions while the manuscript was in preparation and for showing us ref.
\cite{goran} prior to publication. We would also like to thank P. Stevenson
for drawing our attention on some of the works in \cite{fuj}.

\newpage

\vspace{10mm}

\centerline {\LARGE FIGURE CAPTIONS}

\vspace{20mm}

Figure 1. One-loop gap equations for the Higgs' self-energies. Solid
lines correspond to $\chi_{45}$, dashed lines to $H_{24}$ and wavy lines
to gauge fields. The blobs represent insertions of full thermal
propagators.

\vspace{30mm}
Figure 2. Plots of $c_1$ (the solid parabola), $c_2$ (the solid straight
line) and $c_3$ (the dashed line), for $g^2=1/4$ and
$\lambda_{\chi}=71 g^2/135$.

\vspace{30mm}
Figure 3. Plots of $c_1$ (the solid parabola), $c_2$ (the solid straight
line) and $c_3$ (the dashed line), for $g^2=1/16$ and
$\lambda_{\chi}=71 g^2/135$.

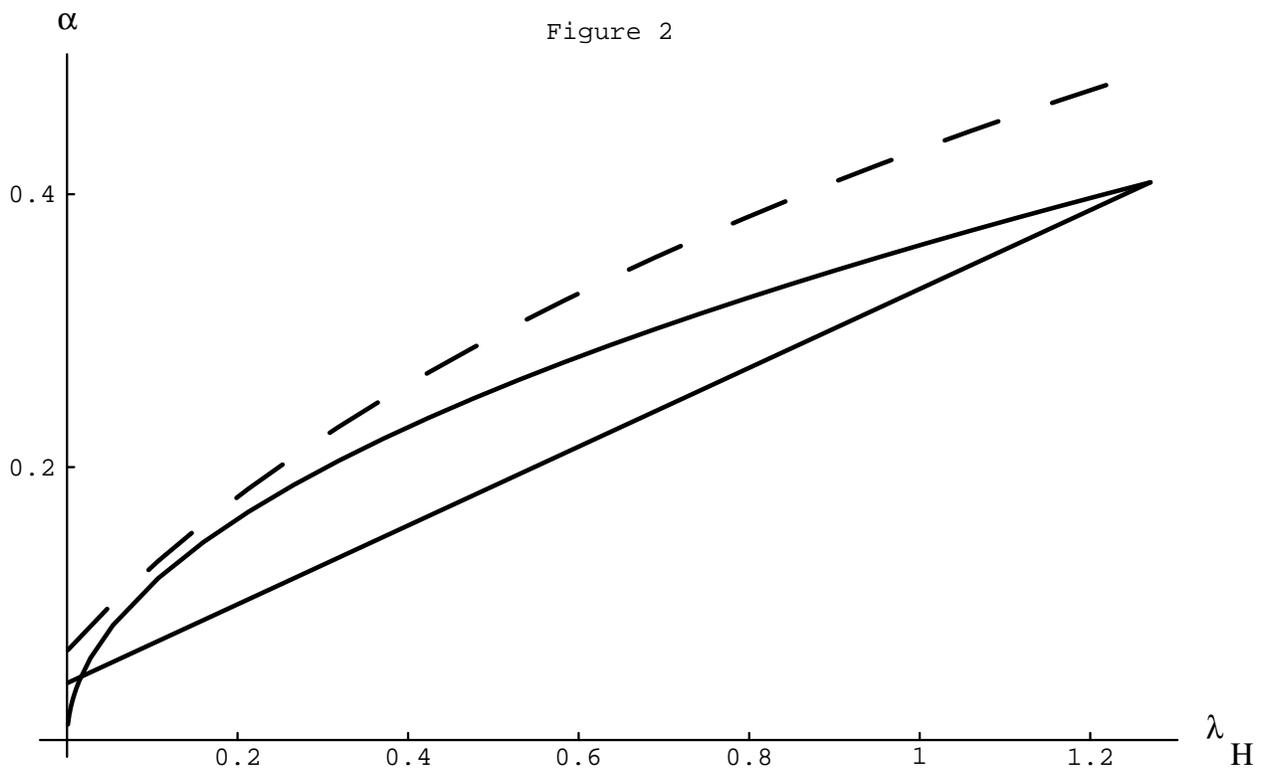
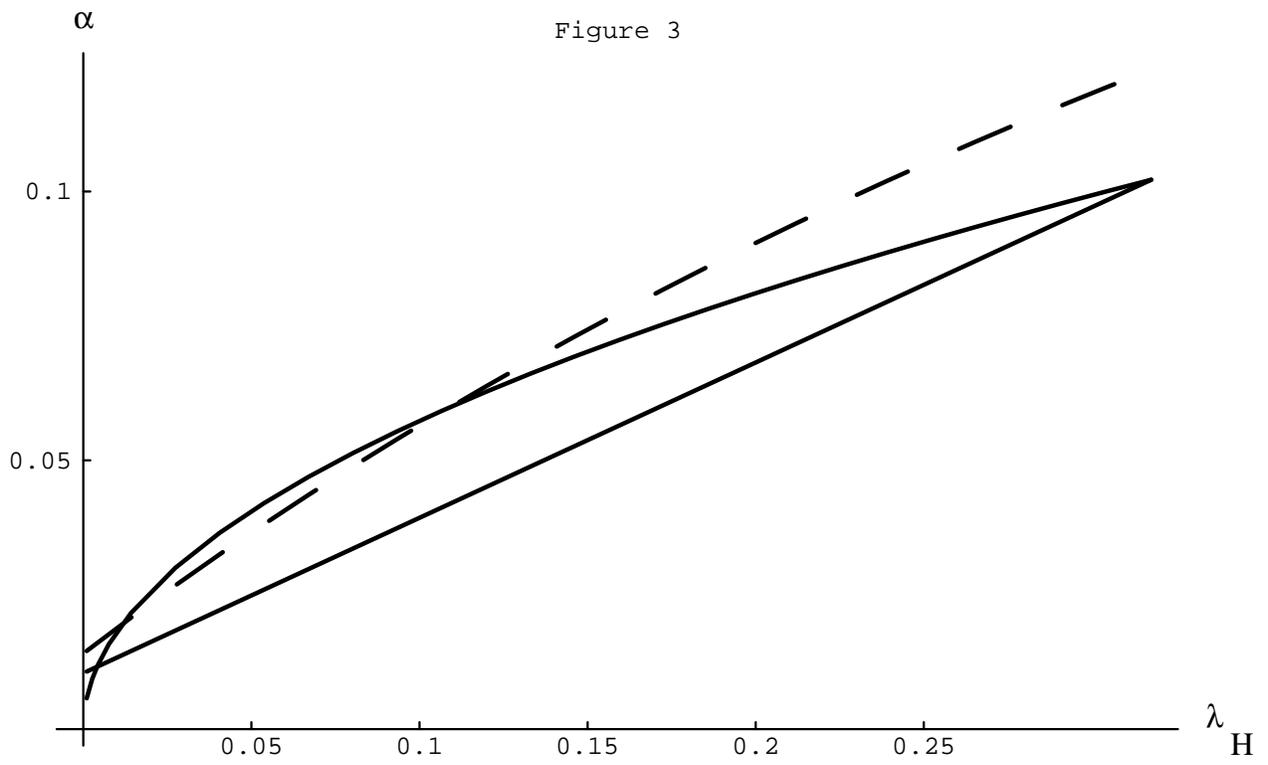
\end{document}